\documentclass[aps,prd,twocolumn,nofootinbib,amsmath,amssymb,showpacs]{revtex4-1}

\usepackage{graphics}
\usepackage[usenames,dvipsnames]{color}
\usepackage{hyperref}

\usepackage{lipsum}

\begin{document}
 
\title{Lepton-rich cold QCD matter in protoneutron stars}

\author{J. C. {\sc Jim\'enez}}
%\email{jjimenez1991@if.ufrj.br}
\author{E. S. {\sc Fraga}}
%\email{fraga@if.ufrj.br}

\affiliation{Instituto de F\'\i sica, Universidade Federal do Rio de Janeiro,
Caixa Postal 68528, 21941-972, Rio de Janeiro, RJ, Brazil}

\date{\today}

%%%%%%%%%%%%%%%%%%%%%%%%%%%%%%%%%%%%%%%%%%%%%%%%%%%%%%%%%%%%%
\begin{abstract}
We investigate protoneutron star matter using the state-of-the-art perturbative equation of state for cold and dense QCD in the presence of a fixed lepton fraction in which both electrons and neutrinos are included. Besides computing the modifications in the equation of state due to the presence of trapped neutrinos, we show that stable strange quark matter has a more restricted parameter space. We also study the possibility of nucleation of unpaired quark matter in the core of protoneutron stars by matching the lepton-rich QCD pressure onto a hadronic equation of state, namely TM1 with trapped neutrinos. Using the inherent dependence of perturbative QCD on the renormalization scale parameter, we provide a measure of the uncertainty in the observables we compute.
\end{abstract}

%%%%%%%%%%%%%%%%%%%%%%%%%%%%%%%%%%%%%%%%%%%%%%%%%%%%%%%%%%%%%

\maketitle

%%%%%%%%%%%%%%%%%%%%%%%%%%%%%%%%%%%%%%%%%%%%%%%%%%%%%%%%%%%%%
\section{Introduction}
%%%%%%%%%%%%%%%%%%%%%%%%%%%%%%%%%%%%%%%%%%%%%%%%%%%%%%%%%%%%%

Neutron stars provide a unique laboratory for the investigation of the strong interaction under extreme conditions \cite{Glendenning:2000}. Especially now, with observations entering a new era: NASA's Neutron star Interior Composition Explorer (NICER) mission \cite{NICER}, which will allow for measurements of neutron star masses and, especially, radii to unprecedented precision, has just become operative, and the first multimessenger observation of a binary neutron star merger has been performed with great success \cite{TheLIGOScientific:2017qsa}, so that gravitational waves can now be used to probe properties of the interior of neutron stars \cite{Andersson:2009yt}.

The crucial ingredient for the description of the structure and phases in the interior of neutron stars is the equation of state (EoS) for neutron star matter, which needs the understanding of the thermodynamics of strong interactions at densities of the order of the saturation density, $n_{0}=0.16~$fm$^{-3}$, and above. Unfortunately, such region in the parameter space of quantum chromodynamics (QCD) is not accessible to a first-principle, nonperturbative (lattice) approach due to the stringent restrictions brought about by the sign problem \cite{deForcrand:2010ys}. The alternative that still provide controlled calculations in the fundamental theory of the strong interactions would be cold and dense perturbative QCD \cite{kapusta-gale,Laine:2016hma}. The state-of-the-art perturbative EoS for cold and dense QCD\footnote{A generalization of the three-loop EoS of zero-temperature quark matter to (small) nonzero temperatures (dubbed cool quark matter) can be found in Ref. \cite{Kurkela:2016was}, while a novel computational aid for perturbative calculations carried out at zero temperature and nonzero density can be found in Ref. \cite{Ghisoiu:2016swa}.} 
was obtained in Ref. \cite{Kurkela:2009gj}, and goes way beyond a simple description based on the MIT bag model as was shown in Ref. \cite{Fraga:2013qra} where the equation of state was also cast into a simple pocket formula assuming local charge neutrality and beta-equilibrium\footnote{It is also possible to build an effective bag model from the two-loop massless cold QCD EoS \cite{Fraga:2001id} and use it to study hybrid stars \cite{Fraga:2001xc,Alford:2004pf}.}. 
Of course, due to asymptotic freedom, this approach is valid only at high enough densities and has to be matched either onto a phenomenological hadronic equation of state at lower densities or to the other controlled limit of QCD, chiral effective field theory (see, e.g. Refs. \cite{Hebeler:2013nza,Tews:2012fj}). One can also use both limits of QCD and a parametrization of the ignorance about the intermediate density region in terms of multiple polytropes to constrain the neutron star matter EoS down to $30 \%$ \cite{Kurkela:2014vha}. Implementing the astrophysical constraints on the maximum mass from measurements of the pulsars PSR J$1614-2230$, with M$=1.97\pm{0.04}$M$_{\odot}$ \cite{Demorest:2010bx}, and PSR J$0348+0432$, with M$=2.01\pm{0.04}$M$_{\odot}$ \cite{Antoniadis:2013pzd}, the previous approach sets limits on some global properties of rotating neutron stars \cite{Gorda:2016uag} besides the mass-radius diagram.

Although cold and dense perturbative QCD (pQCD) can be used to describe the high-density sector of the EoS for neutron stars, it is not adequate to investigate the early stages of their lives as protoneutron stars (PNS), since during the early post-bounce stage of core collapse supernovae matter is still hot and lepton rich. In particular, one has to include the trapped neutrinos in the framework. Reference  \cite{Fraga:2015xha} provides the first attempt of an extension including neutrinos and thermal effects in the case of massless quarks. 

In this paper we investigate protoneutron star matter using the state-of-the-art perturbative equation of state for cold and dense QCD in the presence of a fixed lepton fraction in which both electrons and neutrinos are included. The presence of trapped neutrinos modifies the parameter space of the cold pQCD equation of state. Besides computing the modifications in the equation of state due to the presence of trapped neutrinos, we show that stable strange quark matter is less favorable in this environment. 

Thermal nucleation of a quark phase in supernova matter was previously investigated using a simplified description of the EoS for quark matter in Refs. \cite{Mintz:2009ay,Mintz:2010mh}, where it has been shown that the formation of the first droplet of a quark phase might be very fast and therefore the phase transition to quark matter could play an important role in the mechanism and dynamics of supernova explosions. Here we study the possibility of nucleation of unpaired quark matter in the core of protoneutron stars by matching the lepton-rich QCD pressure onto a hadronic equation of state, namely TM1 with trapped neutrinos. Using the inherent dependence of perturbative QCD on the renormalization scale parameter, we provide a measure of the uncertainty in the observables we compute.

This paper is organized as follows. In Sec. \ref{sec:QM} we discuss the main properties of the equation of state from perturbative QCD in the presence of electrons and trapped neutrinos to describe the lepton-rich thermodynamics. We also analyze the allowed parameter space for stable strange quark matter. In Sec. \ref{sec:phen} we discuss the framework for the description of nucleation of quark matter droplets in protoneutron stars. In Sec. \ref{sec:result} we present our results for the mass-radius relation and nucleation time. Finally, in Sec. \ref{sec:conclusion} we present our summary and outlook.

%%%%%%%%%%%%%%%%%%%%%%%%%%%%%%%%%%%%%%%%%%%%%%%%%%%%%%%%%%%%%
\section{Lepton-rich thermodynamics}
  \label{sec:QM}
%%%%%%%%%%%%%%%%%%%%%%%%%%%%%%%%%%%%%%%%%%%%%%%%%%%%%%%%%%%%%

%%%%%%%%%%%%%%%%%%%%%%%%%%%%%%%
\subsection{Lepton-rich unpaired quark matter}
%%%%%%%%%%%%%%%%%%%%%%%%%%%%%%%

The pressure for cold and dense QCD with massless quarks was first computed to order $\mathcal{O}(\alpha^{2}_{s})$ about four decades ago \cite{Freedman:1976ub,Baluni:1977ms} (see also Ref. \cite{Toimela:1984xy}), and later recomputed with a modern definition of the running coupling constant and used to model the nonideality in the EoS \cite{Fraga:2001id} (see also Ref. \cite{Blaizot:2000fc}). Quark mass effects have been studied in Refs. \cite{Freedman:1977gz,Farhi:1984qu} in the context of quark stars \cite{Itoh:1970uw} and strange matter, respectively. Later, these effects have been described consistently in the $\overline{\rm MS}$ scheme in Ref. \cite{Fraga:2004gz} to $\mathcal{O}(\alpha_{s})$, where it became clear that quark mass effects and their associate renormalization group running can be significant in the physics of neutron stars. The state-of-the-art perturbative EoS for cold quark matter, including renormalization group effects up to $\mathcal{O}(\alpha^{2}_{s})$ in the strange quark mass and strong coupling up to three-loops was obtained by Kurkela $\textit{et al.}$ \cite{Kurkela:2009gj} (see also Ref. \cite{Kurkela:2010yk} for comparison with astrophysical observations). 

A perturbative calculation of the thermodynamic potential necessarily produces an unknown scale, $\bar{\Lambda}$, associated with the subtraction point for renormalization. In the case of zero temperature and mass and nonzero chemical potential, $\bar{\Lambda}$ is proportional to the quark chemical potential, $\mu$. One expects that at higher orders in pQCD this unphysical dependence will diminish. On the other hand, this feature offers  a quantitative way to estimate the contribution of the remaining, undetermined orders, i.e., it provides a measure of the inherent uncertainty in the result. 

\begin{figure}[ht]
	\begin{center}
	\resizebox*{!}{5.5cm}{\includegraphics{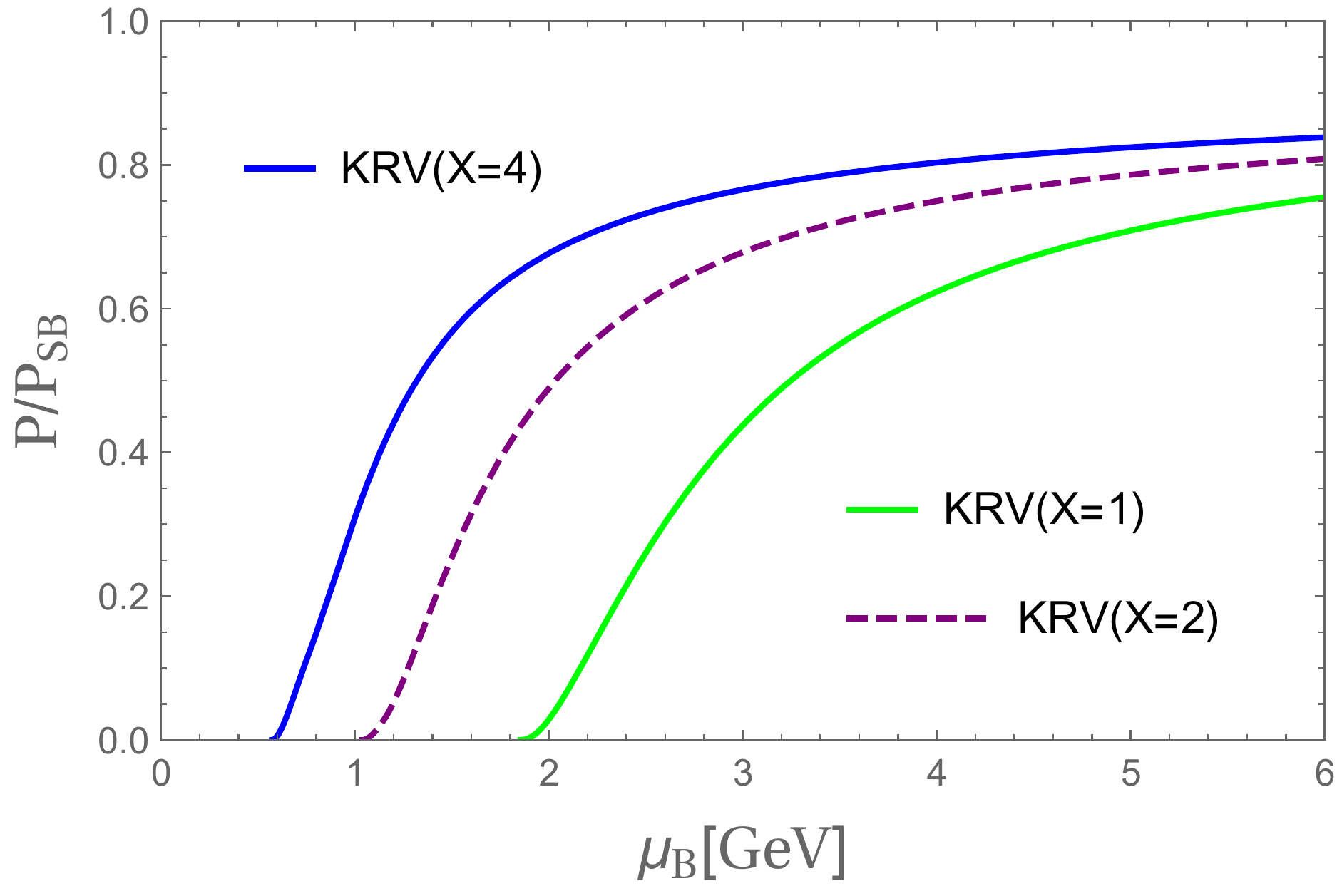}}
      \end{center}    
      \caption{ \label{fig:KRVcold}Total pressure of a gas of up, down and strange quarks plus electrons up to three-loops \cite{Kurkela:2009gj} normalized by the Stefan-Boltzmann free pressure for different values of $X$. Beta equilibrium and local charge neutrality were implemented.}
    \end{figure}

This error band can be estimated by choosing a reasonable fiducial scale and varying $\bar{\Lambda}$ by a factor of $2$. Following Ref. \cite{Kurkela:2009gj}, we adopt the fiducial scale $\bar{\Lambda}=(2/3)\mu_{B}$, where $\mu_{B}$ is the baryon chemical potential\footnote{On phenomenological grounds, one can argue that reasonable values for $\bar{\Lambda}/\mu_q$, where $\mu_q$ is the quark chemical potential, lie between $2$ and $3$, if one takes perturbative QCD as a model for the equation of state for cold strongly interacting matter \cite{Fraga:2001id}.}. 
For the strange quark mass, we choose $m_{s}(2\rm GeV,N _{f}=2+1)=92 \rm MeV$ \cite{Aoki:2016frl} and, for the strong coupling constant, $\alpha_{s}(1.5$GeV$,N_{f}=3)=0.336$, which allows us to fix the renormalization point in the $\overline{\rm MS}$ scheme to $\Lambda_{\overline{\rm MS}}=315^{+18}_{-12}$ MeV \cite{Bazavov:2014soa}. For convenience, we define the dimensionless renormalization scale $X \equiv 3\bar{\Lambda}/\mu_{B}$. The behavior of the pressure obtained in Ref. \cite{Kurkela:2009gj}, which we dub KRV from now on, as a function of $\mu_{B}$ is illustrated in Fig. \ref{fig:KRVcold}.

 %%%%%%%%%%%%%%%%%%%
 %%%%%%%%%%%%%%%%%%%
    
 In order to investigate the early stages of neutron stars lives as protoneutron stars, since during the early postbounce stage of core collapse supernovae matter, one has to consider a lepton-rich EoS. In particular, one has to include the trapped neutrinos in the framework. In what follows we consider a fixed lepton fraction $Y_{i}=n_{i}/n_{B}$, where $n_{B}$ is the baryon density and $n_{i}$ the density of lepton species $i$. So, we take $Y_{L}=Y_{e}+Y_{\nu}\equiv{0.4}$, a value reached in simulations of the evolution of protoneutron stars \cite{Burrows:1986me,Pons:1998mm}.
Associated to this conserved quantity, we introduce an independent neutrino chemical potential $\mu_{\nu}$. 

We start by defining the total quark number density as
\begin{equation}
n=n_{u}(\mu_{u},X)+n_{d}(\mu_{d},X)+n_{s}(\mu_{s},X) \, ,
\end{equation}
where each quark density depends on its respective chemical potential and on the renormalization scale $X$.

Imposing $\textit{local}$ charge neutrality and $\textit{local}$ lepton fraction conservation, we have
      \begin{equation}
%	\label{Beta}
	\frac{2}{3}n_{u}-\frac{1}{3}n_{d}-\frac{1}{3}n_{s}=n^{Q}_{e} \, ,
     \label{neutralYl}
      \end{equation}
      \begin{equation}
%	\label{Beta}
	\frac{n^{Q}_{e}+n^{Q}_{\nu}}{n^{Q}_{B}}=Y_{L}=0.4 \, ,
	\label{fixedYl}     
      \end{equation}
while the weak interaction equilibrium conditions imply
      \begin{equation}
%	\label{Beta}
	\mu_{d}+\mu^{Q}_{\nu}=\mu_{u}+\mu^{Q}_{e},
     \label{eqBetaYl}
      \end{equation}
      \begin{equation}
%	\label{Beta}
	\mu_{d}=\mu_{s}\equiv{\mu} \, .
	\label{strangeYl}     
      \end{equation}
Here, $\mu_{u}$, $\mu_{d}$, $\mu_{s}$, $\mu^{Q}_{e}$ and $\mu^{Q}_{\nu}$ are the chemical potentials of the up, down and strange quarks together with the electron and electron neutrino in the quark phase. The latter are introduced as free Fermi gas contributions. The definition of densities follow straightforwardly. The baryon number density is defined as $n^{Q}_{B}=n/3$. Notice that $\textit{antineutrinos}$, which can play a crucial role in the second neutrino burst signal of a possible QCD phase transition after the first bounce of the core-collapse supernova explosion \cite{Sagert:2008ka}, are also taken into account.

 %%%%%%%%%%%%%%%%%%%
    \begin{figure}[ht]
	\begin{center}
	\resizebox*{!}{5.5cm}{\includegraphics{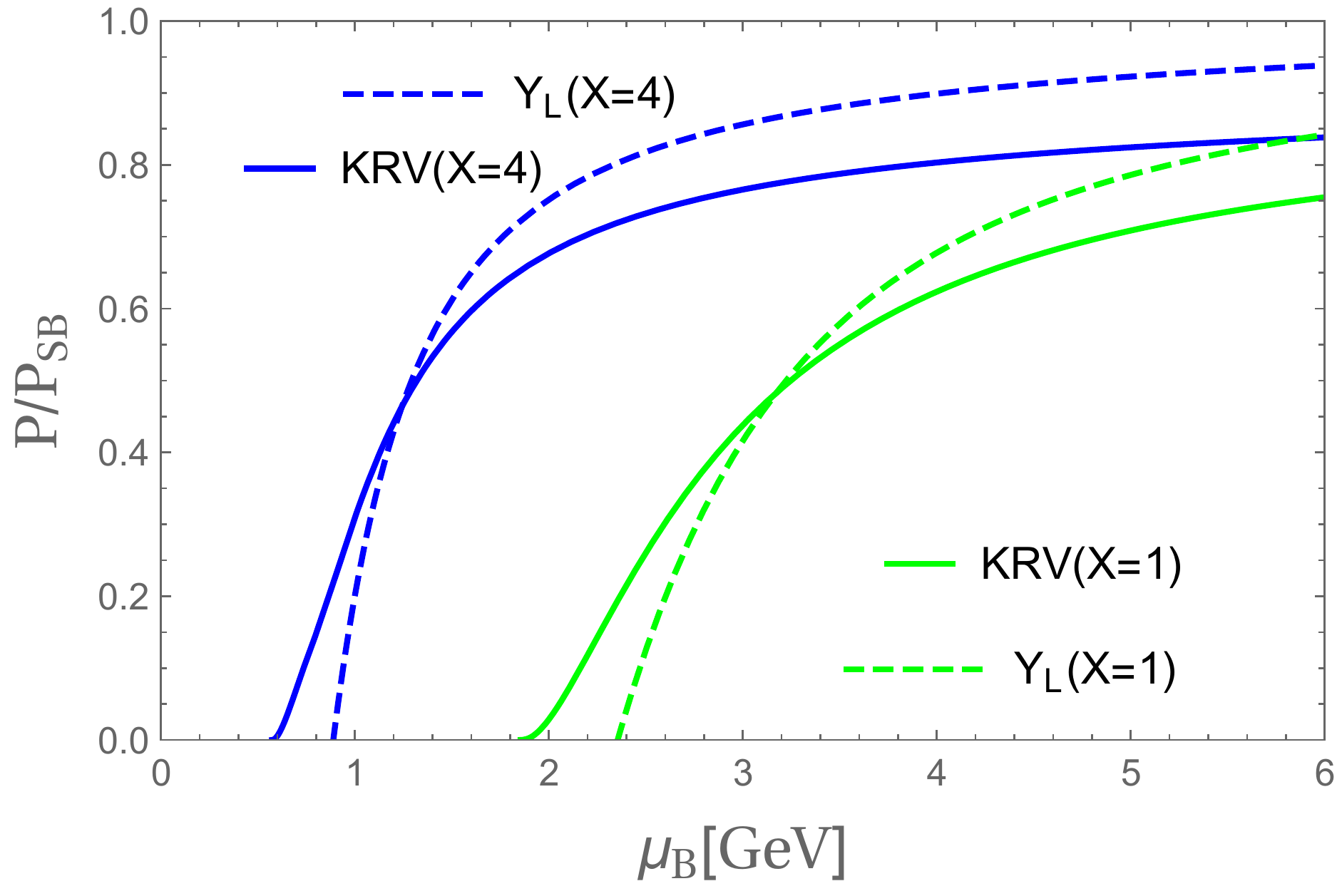}}
      \end{center}    
      \caption{\label{fig:KRVneutrino}Total pressure of quarks and leptons for a fixed lepton fraction ($Y_{L}=0.4$) in dashed lines for the band between $X=1$ and $X=4$. In solid lines we show the lepton-poor case (KRV).}
    \end{figure}
 %%%%%%%%%%%%%%%%%%%

Given the constraints above, we can write all quark and lepton chemical potentials in terms of the strange quark chemical potential, $\mu_{s}\equiv{\mu}$, only. Following Kurkela $\textit{et al.}$ \cite{Kurkela:2009gj}, we use the quark and lepton number densities as the fundamental quantities from which one can construct the pressure, demanding thermodynamic consistency at each step of the calculation and preserving terms up to $\mathcal{O}(\alpha^{2}_{s})$. This procedure makes the implementations of the constraints above on charge neutrality, lepton fraction conservation and chemical equilibrium straightforward. 

\begin{figure}[ht]
	\begin{center}
	\resizebox*{!}{5.5cm}{\includegraphics{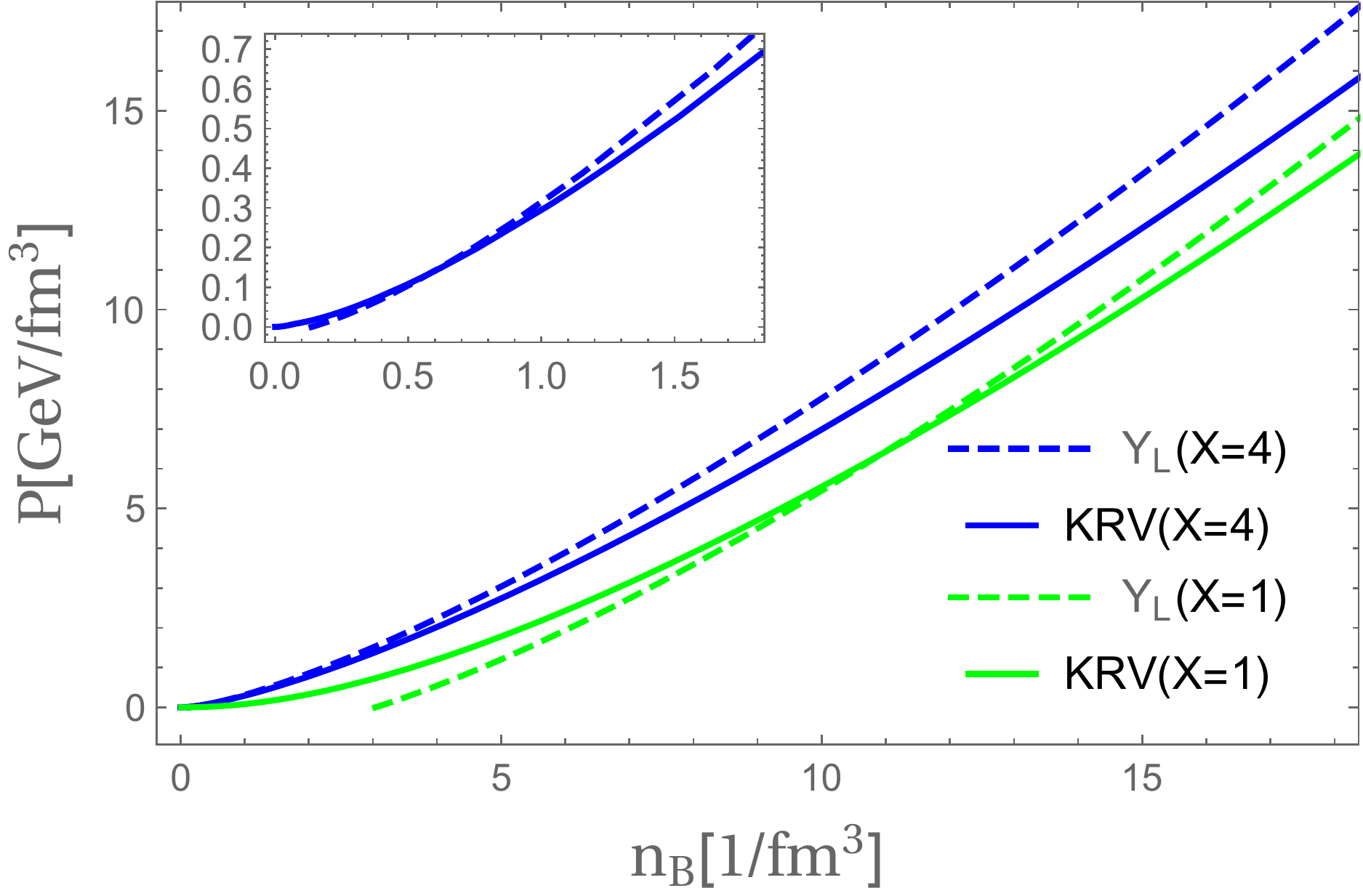}}
      \end{center}    
      \caption{\label{fig:PnBYl} Total pressure of quarks and leptons for a fixed lepton fraction ($Y_{L}(X)$) and in the lepton-poor case, KRV(X) as a function of the baryon number density $n_{B}$.}.
    \end{figure}

In the cases where some quark density $n_{i}(\mu_{i},X)$ becomes negative below a given chemical potential, $\mu_{i}<\mu_i^{0}(X)$, we set it to $n_{i}\equiv{0}$. Integrating the number densities from their minimal value $\mu_i^{0}(X)$ to some arbitrary strange quark chemical potential $\mu$ and taking into account Eqs. (\ref{neutralYl}) --(\ref{strangeYl}), we obtain the total pressure for lepton-rich quark matter as follows:
\begin{eqnarray}
%\begin{aligned}
P(\mu,X)=\int^{\mu}_{\mu_{0}(X)}d\bar{\mu}
\left[ n_{u}\left(1+\frac{d\mu_{\nu}}{d\mu_{s}}-\frac{d\mu_{e}}{d\mu_{s}}\right) \right.\nonumber \\
\left. +~ n_{d}+n_{s}+
n_{e}\frac{d\mu_{e}}{d\mu_{s}}+n_{\nu}\frac{d\mu_{\nu}}{d\mu_{s}} \right] \, .
%\end{aligned}
\end{eqnarray}
We can also express the pressure as a function of the baryon chemical potential, $P=P(\mu_{B})$, where $\mu_{B}=\mu_{u}+\mu_{d}+\mu_{s}$. In Fig. \ref{fig:KRVneutrino} one can see how the cold quark matter EoS (KRV) is modified by the presence of trapped neutrinos ($Y_{L}$) for different values of the renomalization scale $X$. 
Notice that, at high $\mu_{B}$, since $m_{s}(X)$ tends to be constant \cite{Fraga:2004gz,Kurkela:2009gj} and $\alpha_{s}(X)$ is nonzero (unless we are at asymptotically high densities \cite{Kurkela:2009gj}, which are not relevant for the physics of compact stars), the total pressure of quarks and leptons will increase faster, in contrast to the lepton-poor case. However, at low $\mu_{B}$ the lepton-poor total pressure appears to be higher. This occurs due to the respective runnings of $m_{s}$ and $\alpha_{s}$. To clarify this issue, we show, in Fig. \ref{fig:PnBYl},  the total pressure of quarks and leptons as a function of the baryonic number density $n_{B}$. It turns out, as one can see from this figure, that the lepton presence makes our EoS stiffer at high densities.

 %%%%%%%%%%%%%%%%%%%
 %%%%%%%%%%%%%%%%%%%

\begin{figure}[ht]
	\begin{center}
	\resizebox*{!}{5.5cm}{\includegraphics{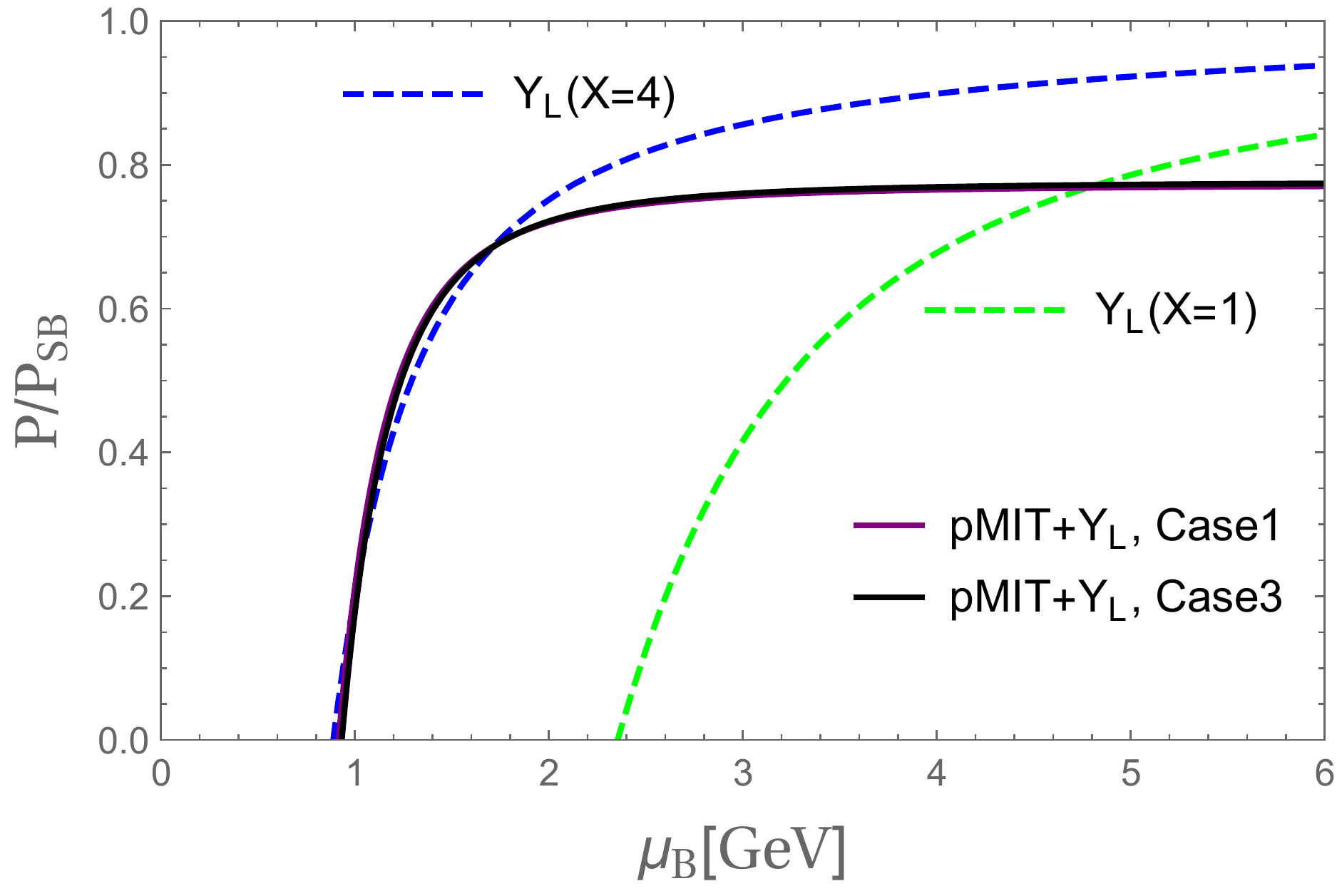}}
      \end{center}    
      \caption{\label{fig:KRVcomparison} Total pressure for a gas of quarks and leptons in beta equilibrium with a fixed lepton fraction $Y_{L}=0.4$ for $X\in[1, 4]$ (dashed lines) in comparison with the effective MIT bag model \cite{Alford:2004pf} which takes into account effectively perturbative corrections for a particular set of values for the strange quark mass, $m_{s}=100 \rm MeV$, and effective bag constants, $B_{1}=(144.65\rm ~ MeV)^{4}$ and $B_{3}=(147.56\rm ~MeV)^{4}$ ($\rm pMIT+Y_{L}$, in solid lines)}.
    \end{figure}

In Fig. \ref{fig:KRVcomparison} we compare our results to the lepton-rich modified version of the effective MIT bag model which takes into account effective corrections from pQCD \cite{Alford:2004pf} (pMIT+$Y_{L}$) for a set of particular values of the effective bag constant $B$ and strange quark mass $m_{s}$ considered in Ref. \cite{Mintz:2009ay}. However, in this reference, the parameters are set to obtain maximum masses of lepton-poor hybrid stars in agreement with observations available then, which were in the order of $1.67$ solar masses, which implies that lower values of the critical baryon density for a deconfinement transition were needed. 

%%%%%%%%%%%%%%%%%%%
 %%%%%%%%%%%%%%%%%%%

%%%%%%%%%%%%%%%%%%%%%%%%%%%%%%%
  \subsection{Lepton-rich stable strange quark matter}
%%%%%%%%%%%%%%%%%%%%%%%%%%%%%%%

Long ago, Bodmer \cite{Bodmer:1971we} and later (independently) Witten \cite{Witten:1984rs} investigated a system formed by massless up, down and strange quarks that, if at zero pressure could have energy per baryon
  \begin{equation}
{E}/{A}\leq{0.93}\rm GeV \, ,
\label{EoverA}
\end{equation}
i.e., lower than the most stable nuclei $\rm Fe^{56}$ (and  $\rm Ni^{62}$), one would find configurations of absolutely $\textit{stable strange quark matter}$ (SQM) as the true ground state of normal matter in the vacuum.

Since then, there have been many improvements trying to simulate the nonperturbative aspects of nuclear strong interactions and taking the mass of the strange quark into account (for reviews, see Refs. \cite{Madsen:1998uh,Weber:2004kj}). Nevertheless, the interactions among quarks are usually described in a very simplified fashion and the presence of neutrinos is overlooked, except in the case of neutrino cooling.

If one is interested in the very first seconds after the formation of stable strange quark matter (either in a cosmological QCD phase transition or in strange stars), one would have to include the presence of neutrinos. In the past it was believed that neutrino cooling of strange matter would be faster than that of neutron star matter \cite{Alcock:1986hz}. Later, this view has been modified by the finding that ordinary neutron beta decay may be energetically allowed also in nuclear matter and so the neutrino cooling could be of the same order as in strange matter \cite{Lattimer:1991ib,Pethick:1991mk,Schaab:1997nt}, a feature that could be investigated in x-ray satellites in the near future.

We can investigate the likelihood of satisfying the criterion given by Eq. (\ref{EoverA}) using the thermodynamic potential we have at hand. As before, it is more convenient to use the quark densities as fundamental quantities and analyze if the parameter space of our theory with strange quark matter is modified in the presence of neutrinos for the usual values of the renormalization scale $X\in[1,4]$. To do this, we use the Hugenholtz-Van Hove theorem \cite{Hugenholtz:1958zz} generalized to a system with many components \cite{R.C.Nayak:2011hyj}. It requires only the quark and lepton densities and chemical potentials as input, giving the following energy per baryon:
	\begin{equation}
	\frac{E(\mu_{s},X)}{A}=\frac{n_{u}}{n^{Q}_{B}}(\mu_{\nu}-\mu_{e})+3\mu_{s} \, ,
	\end{equation}
where we implicitly assumed that all the quantities on the rhs of the equation above are functions of the strange chemical potential $\mu_{s}$.

\begin{figure}[ht]
	\begin{center}
	\resizebox*{!}{5.5cm}{\includegraphics{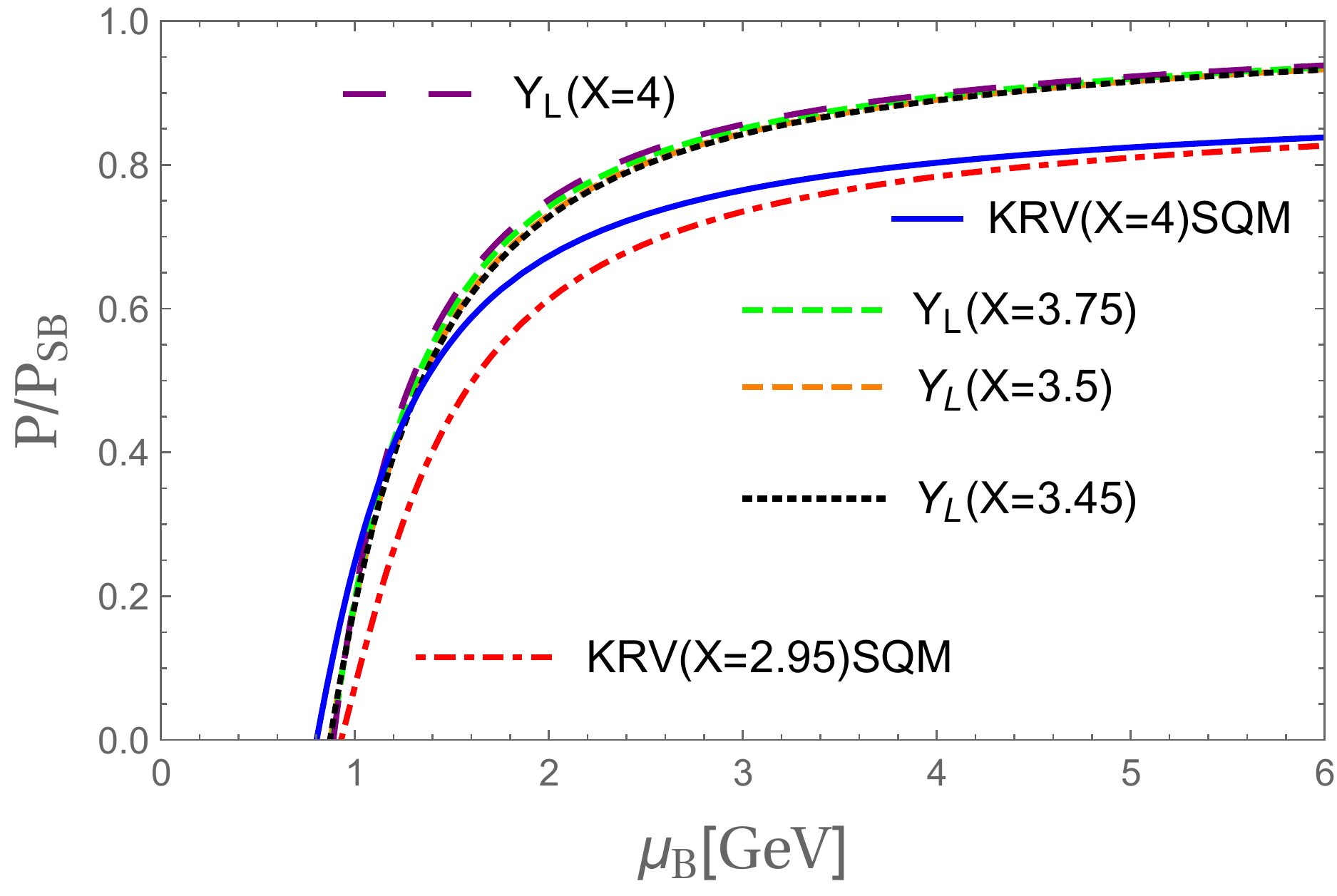}}
      \end{center}    
      \caption{\label{fig:SQMcomparison} Total normalized pressure for quarks and leptons with $Y_{L}=0.4$ allowing for the SQM hypothesis ($Y_{L}$, in dashed lines). For comparison, we show also the pressure for lepton-poor strange matter ($\rm KRV[X]SQM$, in solid and dot-dashed lines).}
    \end{figure}

Constraining the values of $X$ such that $\mu_{s}$ and $n_{s}$ are not zero $\textit{and}$ satisfy Eq. (\ref{EoverA}) we obtain, for the cold case, $X\in[2.95, 4]$, and for the lepton-rich case $X\in[3.45, 4]$, as can be seen in Fig. \ref{fig:SQMcomparison}. Even if the parameter space of $X$ is not radically modified when trapped neutrinos are included, one can notice from Fig. \ref{fig:SQMcomparison} that the band for $X$ tends to shrink to $\mu_{B}\in[0.86, 0.88]\rm GeV$ for vanishing pressure (as compared to $\mu_{B}\in[0.803, 0.93]\rm GeV$ in the cold case). Fig. \ref{fig:SQMcomparison} also indicates that lepton-rich strange quark matter becomes essentially $\textit{independent}$ of the renormalization scale $X$.

 %%%%%%%%%%%%%%%%%%%
 %%%%%%%%%%%%%%%%%%%

In this way one realizes that the presence of neutrinos makes the SQM hypothesis $\textit{less}$ favorable, i.e., the stability windows of critical densities with vanishing pressure is narrower. A similar behavior was observed in Ref. \cite{Dexheimer:2013czv}, where the authors also included finite-temperature contributions in different quark models. This leaves us with $X\in[1, 3.44]$ for unpaired quark matter having as ground state hadronic matter in vacuum.

%%%%%%%%%%%%%%%%%%%%%%%%%%%%%%%%%%%%%%%%%%%%%%%%%%%%%%%%%%%%%
\section{Nucleating quark matter in PNS}
  \label{sec:phen}
%%%%%%%%%%%%%%%%%%%%%%%%%%%%%%%%%%%%%%%%%%%%%%%%%%%%%%%%%%%%%
  
From now on we assume that strange quark matter is not the true ground state of strong interactions and consider the nucleation of quark matter droplets in the core of a protoneutron star (PNS) which, after a brief period of deleptonization, will produce a regular neutron star. In particular, we investigate whether unpaired quark matter can be nucleated as the density increases in a medium of hadronic matter rich in trapped neutrinos.
In the case of a first-order phase transition in which we still have the presence of a barrier to fluctuations, it is well known that there are two main mechanisms for the nucleation of the true ground-state phase: thermal activation and quantum tunneling \cite{Gunton:1983}. A few years ago, it has been shown that, in PNS conditions, thermal activation will dominates over quantum nucleation \cite{Mintz:2009ay}. So, our analysis will focus on the thermal nucleation of quark matter droplets in a medium of hadronic matter. 

In order to do that, we have to match the lepton-rich pQCD EoS discussed in the previous section onto an EoS for the hadronic phase. Given the difference in scale between the temperatures involved in PNS matter ( $T\sim 20~$ MeV) and the typical baryon chemical potential in the medium, we can keep cold ($T=0$) approximations for the EoS. Thermal corrections would be of the order of $\mathcal{O}(T^{2}/\mu^{2}_{B})\sim{1}\%$. Therefore, temperature effects will matter only in the calculation of the nucleation rate, as implemented, e.g. in Refs. \cite{Mintz:2009ay,Palhares:2010be}.

%%%%%%%%%%%%%%%%%%%
\subsection{Matching onto a hadronic EoS}
%%%%%%%%%%%%%%%%%%%

There are many equations of state that can be used to describe the properties of lepton-poor cold nuclear matter at densities around the nuclear saturation density, $n_{0}$. Some of them describe correctly the many phases that can exist inside a (cold) neutron star. However, for PNS matter the most appropriate (and usual) choice of EoS is the one which comes from relativistic mean field theory using the so-called TM1 parametrization of Shen $\textit{et al.}$ \cite{Shen:1998gq}. Here, we had to generalize their lepton-poor result to the case where neutrinos are trapped. Notice that this stiff EoS softens when we add a fixed fraction of leptons, in contrast to our lepton rich QCD EoS.
    
Hence, we impose again local charge neutrality, $n_{p}=n^{H}_{e}$, local lepton fraction conservation (cf. Ref.\cite{Pagliara:2009dg} for a global version of this constraint in PNS matter)
      \begin{equation}
	\label{Beta}
	\frac{n^{H}_{e}+n^{H}_{\nu}}{n^{H}_{B}}=Y_{L}=0.4,
      \end{equation}
and the weak equilibrium condition $\mu_{n}+\mu^{H}_{\nu}=\mu_{p}+\mu^{H}_{e}$, where $\mu_{n}$, $\mu_{p}$, $\mu^{H}_{e}$ and $\mu^{H}_{\nu}$ are the chemical potentials of neutrons, protons, electrons and electron neutrinos in the hadronic phase, respectively. Also, $n_{p}$, $n_{n}$, $n^{H}_{e}$ and $n^{H}_{\nu}$ are the respective particle densities with $n^{H}_{B}=n_{n}+n_{p}$, being the baryon number density. Once more, lepton densities are introduced as free Fermi gases. When we refer to this lepton-rich hadronic matter EoS we will use the abbreviation TM1-PNS.

The scenario we have in mind is that of the core-collapse of a supernova, similar to the one studied in Refs. \cite{Mintz:2009ay,Palhares:2010be}. Not taking into account the strange quark matter hypothesis, we know that for some density region there could be a deconfinement phase transition between hadronic and quark matter phases in a very dense and neutrino-rich environment found in PNS matter \cite{Glendenning:2000}. Since initially the hadronic phase does not contain strangeness, one would expect weak interactions could trigger a phase transition to unpaired quark matter. However, this would be too slow to produce strange quarks compared to the fast deconfinement transition driven by strong interactions \cite{Norsen:2002qw,Mintz:2009ay}.

Another scenario would be to consider a fast production of strange quarks due to the environment conditions of temperature and density of PNS matter, where a small amount of strangeness may appear through the presence of hyperons \cite{Bhattacharyya:2006vy}. Hyperons could then convert two-flavor quark matter into unpaired quark matter. Although it is well known that the presence of neutrinos inhibits the presence of hyperons at high densities, statistical fluctuations can be important \cite{Ishizuka:2008gr,Pons:1998mm}. The scenario adopted in this work is somewhat more inclusive since our EoS naturally adds a strange massive component to the two-flavor quark matter EoS as one goes from low to high densities, so that this scenario unifies the ones above for the formation of unpaired quark matter.

From the neutrino-rich pQCD EoS, the phase transition could be of first-order, depending on the chosen value for the renormalization scale $X$. In that case, we can use the modified Maxwell construction of Ref. \cite{Hempel:2009vp} for PNS matter, which mimics the out-of-equilibrium conditions. Then, our conditions for phase coexistence are the equality of the total pressures of the two phases, $P^{H}=P^{Q}$, and the condition of chemical equilibrium
  \begin{equation}
  \mu_{n}+Y_{L}\mu^{H}_{\nu}=\mu_{u}+2\mu_{d}+Y_{L}\mu^{Q}_{\nu}=\mu_{B}\equiv
  \mu_{\rm eff} \, .
  \label{mu_eff}
  \end{equation}

Indeed, although $P^{H}=P^{Q}$ is valid only at the transition point, Eq. (\ref{mu_eff}) is the chemical potential associated to the $\textit{global}$ conservation of baryon number along all the PNS life. To avoid confusion with the cold lepton-poor case, where $\mu_{n}=\mu_{B}$ for the hadron phase and $\mu_{u}+2\mu_{d}=\mu_{B}$ for the quark phase, we define it as an effective chemical potential, $\mu_{\rm eff}$, valid through all the phases of the PNS and useful for our matching procedure.

Since we assume a first-order phase transition, it is appropriate to use a thermodynamic quantity which tells us how $\textit{strong}$ is the phase transition. This quantity will be the latent heat, defined as $\Delta{Q}{\equiv}\mu^{c}_{\rm eff}\Delta{n_{B}}$\cite{Kurkela:2014vha}, where $\mu^{c}_{\rm eff}$ is the critical effective chemical potential from which the quark phase starts and $\Delta{n_{B}}$ the baryonic number difference between phases at the critical point\footnote{In principle, a correct matching for both phases of lepton-rich matter would imply the existence of a mixed phase with possible nontrivial geometrical structures \cite{Glendenning:2000}. However, modifying our conditions of charge neutrality from local to global would not affect considerably our results in comparison to the intrinsic uncertainty brought about by the renormalization scale dependence of our results.}.

%%%%%%%%%%%%%%%%%%%
\subsection{Nucleation time}
%%%%%%%%%%%%%%%%%%%

To estimate the time scale for nucleation of cold deconfined quark matter with a fixed  lepton fraction $Y_{L}$, we need to compute the nucleation rate, $\Gamma$. The standard Langer formalism for homogeneous nucleation via thermal activation yields, in the thin-wall limit \cite{Langer:1969bc} (see also Ref. \cite{Gunton:1983})
\begin{equation}
\Gamma=\frac{\mathcal{P}_{0}}{2\pi}\exp\left[-\frac{\Delta{F}(R_{c})}{T}\right] \, ,
\end{equation}
where $\Delta{F}(R_{c})$ is the difference in free energy between the metastable (nuclear) phase and the true stable (quark) phase, which can be written in terms of the radius of the critical bubble, $R_{c}$, which corresponds to a saddle point in functional space. As customary, we can take the upper limit estimate for the pre-factor ${\mathcal{P}_{0}}/{2\pi}=T^{4}$ as \cite{Csernai:1992tj}. We can rewrite $\Gamma$, after some straightforward algebra, as
   \begin{equation}
   \Gamma=T^{4}\exp\left[-\frac{16\pi}{3}\frac{{\sigma}^{3}}{(\Delta{P})^{2}T}\right],
   \label{eq:gamma}
   \end{equation}
where $\Delta{P}$ is the difference between the matched pressures. It is clear that the surface tension, $\sigma$, plays a crucial role \cite{Mintz:2009ay,Palhares:2010be,Pinto:2012aq,Mintz:2012mz,Lugones:2013ema}.
  
Hence, the nucleation time, $\tau$, to create the first single {\it critical} droplet of lepton-rich unpaired quark matter inside a volume of $1 \rm km^{3}$, which is the typical size of the core of a PNS, is given by \cite{Mintz:2009ay}
\begin{equation}
\tau_{\rm nucl}\equiv\left({\frac{1}{1 \rm km^{3}}}\right)\frac{1}{\Gamma} \, ,
\label{eq:tau_nucl}
\end{equation}
where we assume homogeneity of density and temperature in the core, a good approximation since the density profile in this region of the PNS is approximately flat \cite{Glendenning:2000}.

%%%%%%%%%%%%%%%%%%%%%%%%%%%%%%%%%%%%%%%%%%%%%%%%%%%%%%%%%%%%%
\section{Results}
	\label{sec:result}
%%%%%%%%%%%%%%%%%%%%%%%%%%%%%%%%%%%%%%%%%%%%%%%%%%%%%%%%%%%%%

%%%%%%%%%%%%%%%%%%%	
\subsection{Hybrid PNS}
%%%%%%%%%%%%%%%%%%%
  
To be consistent with current astrophysical observations of neutron star masses allowing for a core of quark matter \cite{Alford:2006vz}, one has to choose values for $X$ in the cold lepton-poor pQCD equation of state, KRV-EoS, that match the cold lepton-poor TM1-EoS\footnote{For the low-density region, we include the Baym-Pethick-Sutherland EoS \cite{Baym:1971pw} that is necessary for an adequate treatment of the crust.} 
at a given critical baryon chemical potential and generate at least two-solar mass stars as maximum masses. Solving the Tolman--Oppenheimer-Volkoff (TOV) equations of hydrostatic equilibrium \cite{Glendenning:2000} for the matched EoSs, we find that both $\textit{soft}$ and $\textit{strong}$ first-order phase transitions are able to accommodate masses above the latest measurements for pulsars in binary systems.

	\begin{figure}[ht]
	\begin{center}
	\resizebox*{!}{5.5cm}{\includegraphics{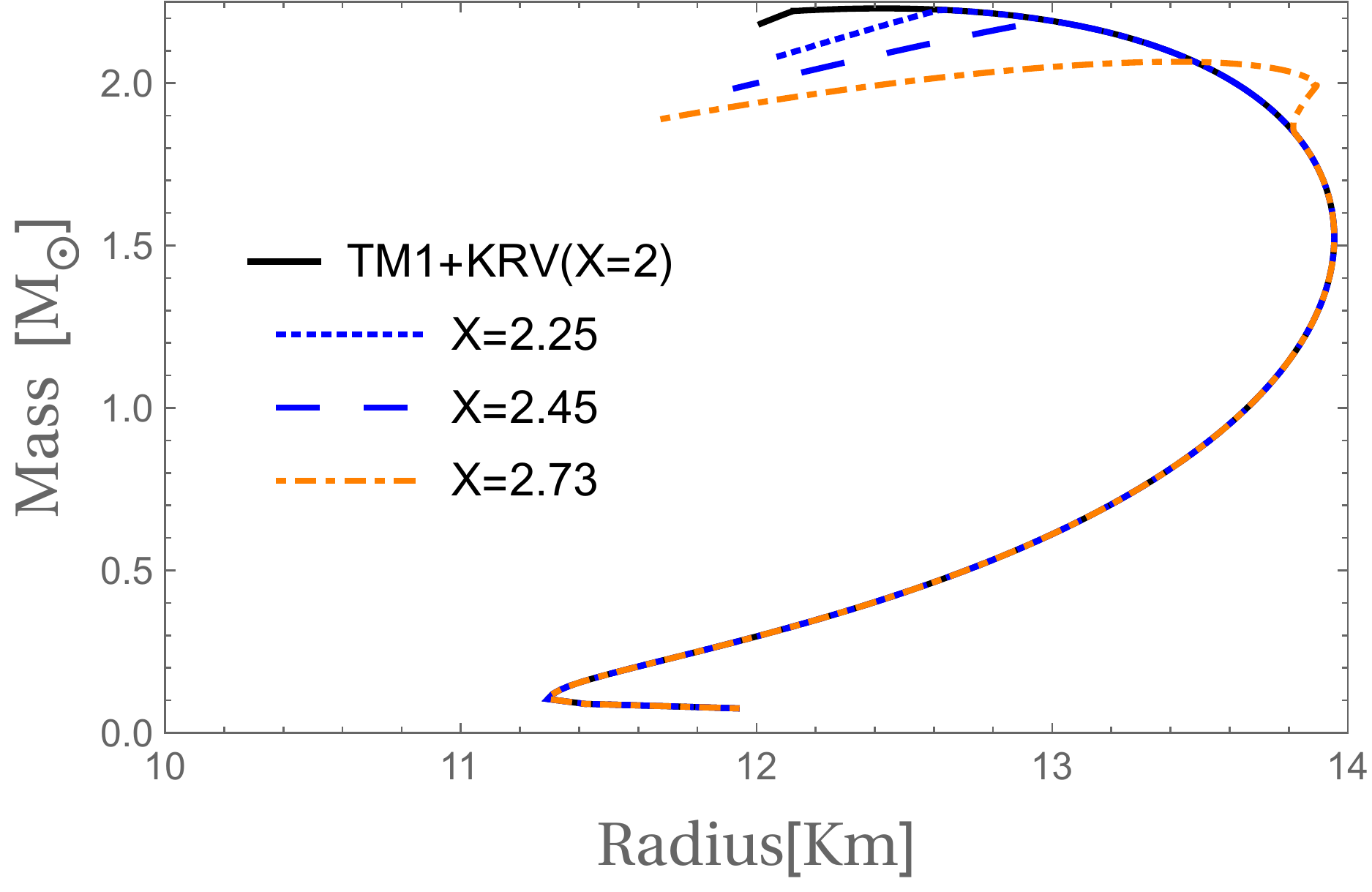}}
      \end{center}    
      \caption{\label{fig:MR} Mass-radius diagram for hybrid stars that masquerade (pure or mixed phase) quark matter cores. The matching is performed using the TM1-EoS for the hadronic phase and the KRV-EoS for the quark phase for different values of the renormalization scale $X$.}
    \end{figure}

In Fig. \ref{fig:MR} we show the mass-radius diagram for a few values of the renormalization scale $X$ featuring first-order phase transitions with different critical densities and intensities. We restrict the values of the renormalization scale to the interval $[2,2.73]$--- if one keeps higher values, one would produce hybrid stars that do not satisfy the observational constraint of two solar masses; if one keeps lower values of $X$, one would find purely nucleonic stars.

The softest possible matching still exhibiting a first-order phase transition to a quark mixed phase at a critical baryon density of $n_{\rm crit}=2.82n_{0}$, with a pure quark core with central density $n_{c}=4.2n_{0}$, corresponds to $X=2.73$. The maximum mass in this case can reach $M=2.08M_{\odot}$, and the latent heat is given by $\Delta{Q}(\rm X=2.73)=(129.3\rm MeV)^{4}$, the lowest value of latent heat obtained in the matching procedure. Following the arguments of Ref. \cite{Kurkela:2014vha}, one expects to have soft first-order phase transitions if the latent heat $\Delta{Q}$ is smaller than $(\Lambda_{\rm QCD})^{4}$, so that one has a large parameter space to surpass the two-solar mass limit with a quark content in hybrid stars.

As we decrease the value of $X$, the first-order phase transition becomes stronger and happens at very high densities, as can be seen in Table I, making the hybrid star more nucleonic with a small mixed quark core. 

\begin{table}[h!]
  \begin{center}
    \label{tab:table1}
    \begin{tabular}{c|c|c} % <-- Changed to S here.
      $X$ & $n_{\rm crit}$ & $\Delta{Q}$\\
      \hline
      $2$ & $9n_{0}$ & $(286.5\rm ~MeV)^{4}$\\
      $2.25$ & $6.85n_{0}$ & $(251.9\rm ~MeV)^{4}$\\
      $2.45$ & $5.35n_{0}$ & $(221.4\rm ~MeV)^{4}$\\
    \end{tabular}
        \caption{Table of critical baryon densities and latent heats for different values of $X$.}
  \end{center}
\end{table}

\begin{figure}[ht]
	\begin{center}
	\resizebox*{!}{5.5cm}{\includegraphics{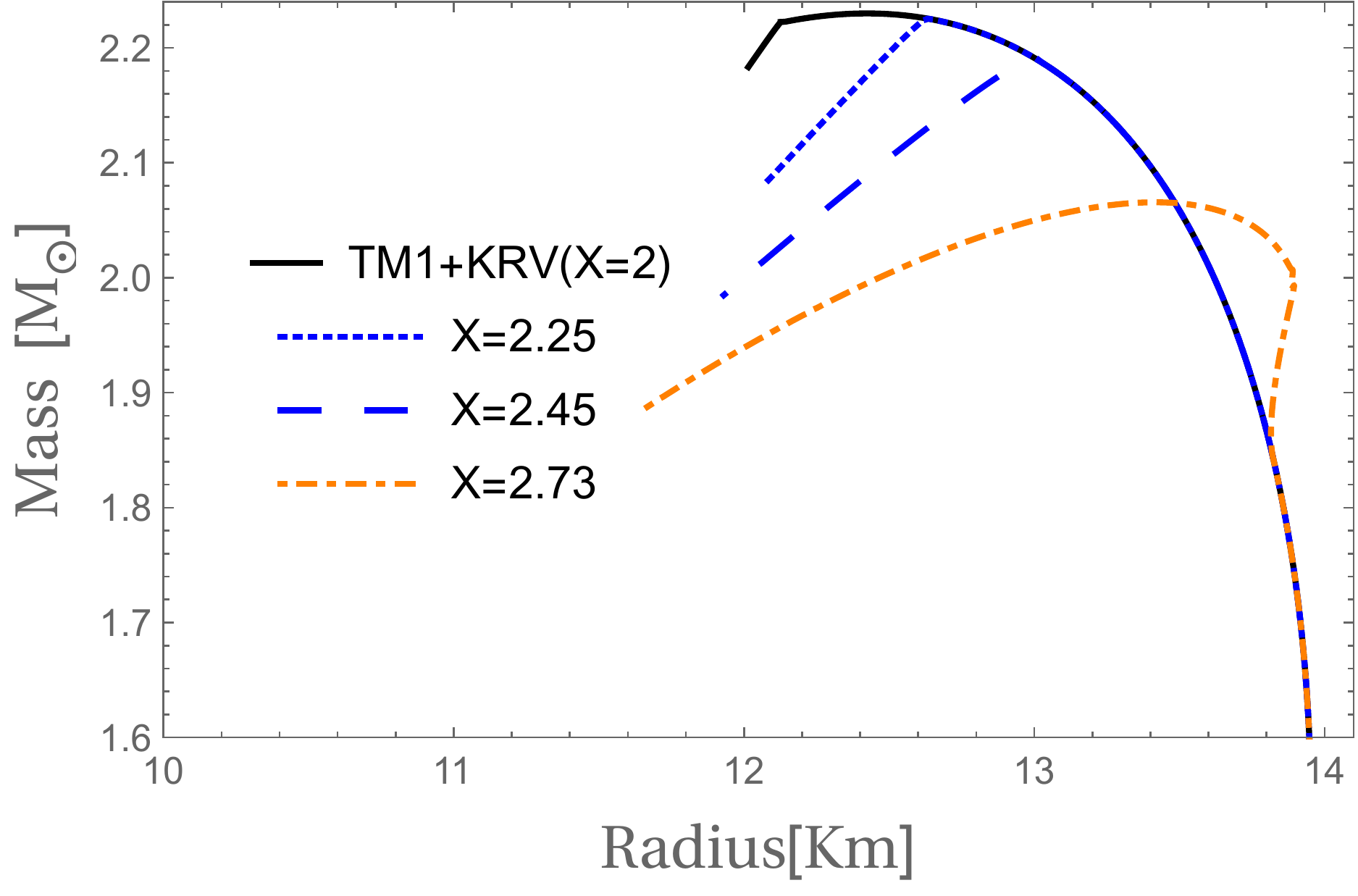}}
      \end{center}    
      \caption{\label{fig:zMR}Zoom in the maximum mass region in our M-R diagram.}
    \end{figure}

Solving the TOV equations for matched pressures with $\rm X< 2.73$ gives us different sequences of stellar configurations. In Fig. \ref{fig:zMR} we show a zoom in the mass-radius diagram for the region related to the maximum mass limits. It is clear that, in order to achieve maximum masses above $2 \rm M_{\odot}$ and satisfy the observational constraints on the cold deleptonized hybrid star, we should use for the lepton-rich EoS values in the band $X\in[2, 2.73]$. The curves in Fig. \ref{fig:zMR}  show the usual behavior for hybrid stars built using the Maxwell construction\footnote{This construction is justified by the fact that we are assuming implicitly two homogeneous locally neutral phases and high enough values for the (unknown) surface tension.} 
(see Ref. \cite{Glendenning:2000}), including the sharp breaking of the curves other than the case $X=2.73$ (in which the homogeneous phase of constant pressure is very small). One should be aware that the details of the construction are, of course, model dependent.

%      

%%%%%%%%%%%%%%%%%%%
\subsection{Nucleation to unpaired quark matter}
%%%%%%%%%%%%%%%%%%%

We will turn our attention back to the early post-bounce state of core-collapse supernovae producing  protoneutron star (PNS) matter and the question of timescales for the nucleation of quark matter in the core of a PNS. In this situation, neutrinos are explicitly trapped, and one should use the complete lepton-rich pQCD EoS satisfying the constraints mentioned above.

\begin{figure}[ht]
	\begin{center}
	\resizebox*{!}{5.5cm}{\includegraphics{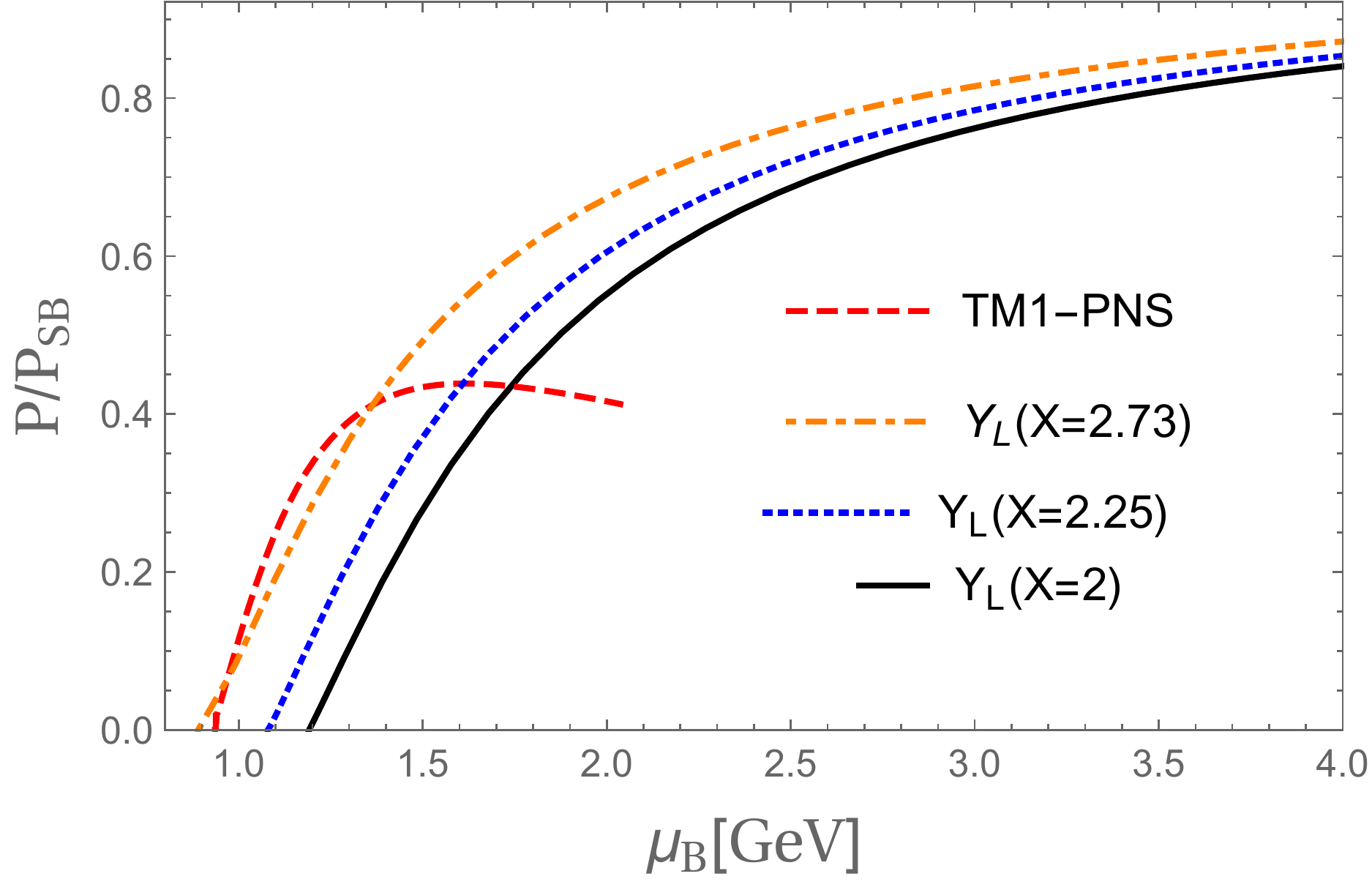}}
      \end{center}    
      \caption{\label{fig:normP} Total normalized pressure for lepton-rich quark matter matched onto lepton-rich hadronic EoS TM1-PNS (dashed line) for different values of $X$ that allow for nucleation of unpaired quark matter still consistent with measurements of two-solar mass pulsars.}
    \end{figure}

In Fig. \ref{fig:normP} we illustrate the matching of a few cases of the lepton-rich pQCD EoS with $X\in[2, 2.73]$ onto the TM1-PNS EoS. One can see that, for $X=2.73$, the phase transition is $\textit{not}$ soft anymore, in contrast to the lepton-poor case. Nevertheless, it occurs at a critical density which is still not very high. Something analogous happens for the other values of the renormalization scale $X$ displayed, so that one can conclude  that the presence of neutrinos shifts the critical densities towards larger values, turning $\textit{weak}$ first-order transitions (in deleptonized dense matter) into $\textit{strong}$ first-order transition in the lepton-rich case.

\begin{figure}[ht]
	\begin{center}
	\resizebox*{!}{5.5cm}{\includegraphics{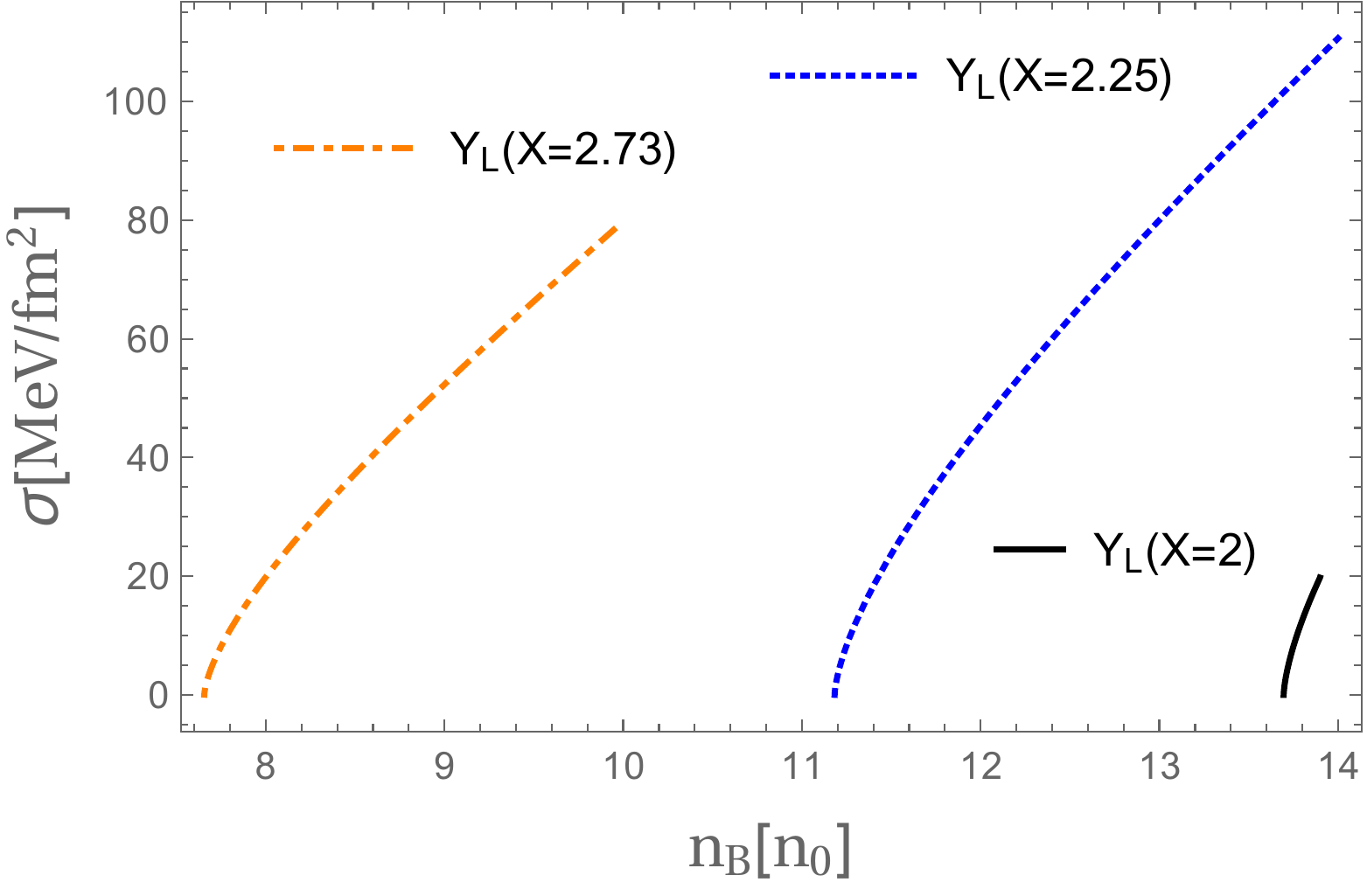}}
      \end{center}    
      \caption{\label{fig:surfacetension} Lines of constant nucleation time $\tau_{\rm nucl}=\tau_{\rm PNS}=100 \rm ~ms$ for a PNS in terms of the surface tension and the baryon density for different values of $X$.}
\end{figure}

For nucleation to be effective, its typical time scale should be of the order of the lifetime of the PNS matter, i.e., $\tau_{\rm nucl}=\tau_{\rm PNS}=100 \rm ~ms$. Then, following the procedure of Ref. \cite{Mintz:2009ay}, we can make a contour plot for different values of surface tension, $\sigma$, and baryon density, using Eqs. (\ref{eq:gamma}) and (\ref{eq:tau_nucl}). These results are shown in Fig. \ref{fig:surfacetension} for three different values for the renormalization scale, namely  $X=2$, $2.25$ and $2.73$, which gives us qualitatively similar behaviors for the rising of surface tension with baryon density for different values of the critical baryon density. 


As we go to higher values of $n_{B}$, there is a small window for nucleation of unpaired quark matter droplets, requiring lower and lower values of the surface tension. The corresponding values for the critical baryon density to form droplets of quark matter and latent heat released at the transition are shown in Table II. Notice that the presence of trapped neutrinos makes the critical densities noticeably higher (cf. Table I), which is in line with previous results of Ref. \cite{Lugones:1997gg}. The latent heat $\Delta{Q^{\nu}}$ is related to the second neutrino burst to be found in supernovae explosions in the case of a QCD transition, as proposed in Ref. \cite{Sagert:2008ka}. Moreover, the difference between latent heats in the lepton-rich and deleptonized cases, $\Delta{Q^{\nu}}-\Delta{Q}$, can be interpreted as the energy taken away in the form of neutrino emission during the deleptonization phase.

\begin{table}[h!]
  \begin{center}
    \label{tab:table2}
    \begin{tabular}{c|c|c} % <-- Changed to S here.
      $X$ & $n_{\rm crit}$ & $\Delta{Q^{\nu}}$\\
      \hline
      $2$ & $13.7 n_{0}$ & $(325.7\rm ~MeV)^{4}$\\
      $2.25$ & $11.2 n_{0}$ & $(297.4\rm ~MeV)^{4}$\\
      $2.73$ & $7.66 n_{0}$ & $(246.1\rm ~MeV)^{4}$\\
    \end{tabular}
        \caption{Table of critical baryon densities and latent heats for different values of $X$ in the lepton-rich case.}
  \end{center}
\end{table}
%

%%%%%%%%%%%%%%%%%%%%%%%%%%%%%%%%%%%%%%%%%%%%%%%%%%%%%%%%%%%%%
\section{Summary and outlook}
  \label{sec:conclusion}
%%%%%%%%%%%%%%%%%%%%%%%%%%%%%%%%%%%%%%%%%%%%%%%%%%%%%%%%%%%%%

In this work we have investigated protoneutron star matter using the state-of-the-art perturbative equation of state for cold and dense QCD in the presence of a fixed lepton fraction in which both electrons and neutrinos are included. Finite-temperature effects can be neglected since they have a minor effect in the PNS scenario at hand. Even if the presence of neutrinos does not modify appreciably the EoS at low densities, their presence significantly increases the pressure as one goes to higher densities, within the region that is relevant for the physics of PNS.

Besides computing the modifications in the equation of state due to the presence of trapped neutrinos, we have shown that stable strange quark matter is less favorable in this environment, i.e., the parameter space for the formation of strange quark matter with neutrinos decreases. 

In order to estimate the odds of nucleating unpaired quark matter in the core of protoneutron stars, we had to match the lepton-rich QCD pressure onto a hadronic equation of state, namely TM1 with trapped neutrinos.

In doing so, we found that neutrinos make the deconfinement transition from nuclear matter to quark matter more $\textit{difficult}$, in line with previous results that use simplified models for the quark matter sector \cite{Lugones:1997gg,Benvenuto:1999uk,Menezes:2003pa}. However, whereas the previous descriptions require very high values of the  critical density, $n_{\rm crit} \sim 10 n_{0}$, so that the maximum masses of PNS surpass the current limit of two solar masses, the description using lepton-rich pQCD needs a critical density $n_{\rm crit} \sim 3 n_{0}$ to produce deleptonized stable hybrid stars satisfying the observational constraints.

Although our results are sensitive to the inherent dependence of perturbative QCD on the renormalization scale parameter, this, on the other hand, provide a measure of the uncertainty in the observables computed.

In this paper, we did not address the protoneutron star evolution taking into account transport properties \cite{Pons:1998mm}. A future study would include the protoneutron star evolution taking into account the transport properties, like what was done in Refs. \cite{Pons:1998mm,Pons:2001ar}, but using our EoS. The deleptonization process also produces a compression (and heating) of the PNS matter, opening a window for instability to gravitational collapse depending on the initial mass \cite{Prakash:1995uw}. This evolution could also be studied using the equation of state provided by the lepton-rich pQCD framework.

%%%%%%%%%%%%%%%%%%%%%%%%%%%%%%%%%%%%%%%%%%%%%%%%%%%%%%%%%%%%%
\begin{acknowledgments}
This work was supported by CNPq, FAPERJ, being also part of the project INCT-FNA Process No. 464898/2014-5.
\end{acknowledgments}

%%%%%%%%%%%%%%%%%%%%%%%%%%%%%%%%%%%%%%%%%%%%%%%%%%%%%%%%%%%%%

\end{document}